%Paper: hep-ph/9210262
%From: hoyer@fltxa.helsinki.fi (Paul Hoyer)
%Date: Mon, 26 Oct 92 11:17:22 EET

\input phyzzx.tex
%macropackage=phyzzx

\overfullrule 0pt
 \PHYSREV
\def\M{{\cal M}}

\def\etal{{\it et. al.}}

 \doublespace
 \pubnum{5935}
 \date{October 1992}
 \pubtype{T/E}
 \titlepage
\vfill
 \title{%
A Bound on the Energy Loss of Partons in Nuclei
 \doeack}
 \author{Stanley J. Brodsky}
 \SLAC
\andauthor{Paul Hoyer}
\address{Department of Physics \break
University of Helsinki, Helsinki, Finland}
\vfill
\abstract

We derive a quantum mechanical upper bound on the amount of radiative
energy loss suffered by high energy quarks and gluons in nuclear matter.
The bound shows that the nuclear suppression observed in quarkonium
production at high $x_F$ cannot be explained in terms of energy loss of
the initial or final parton states.  We also argue that no nuclear
suppression is expected in the photoproduction of light hadrons at large
$x_F$.
\vfill
\submit{Physics Letters}
\vfill

\REF\bart{D.S.  Barton \etal, Phys.  Rev. \us{D27}, 2580 (1983)}

\REF\GW{ M. L. Good and R. Walker, Phys. Rev.
\us {120}, 1855, 1857 (1960). E. L. Feinberg and I. Ia. Pomeranchuk,
Nuovo Cimento Suppl. \us {III}, 652 (1956).}

\REF\coh{G.  Bertsch, S.  J.  Brodsky, A.  S.  Goldhaber, J. F.  Gunion,
Phys.  Rev.  Lett. \us {47}, 297 (1981).  H.
Heiselberg, G.  Baym, B.  Blaettel, L.L.  Frankfurt, M.  Strikman, Phys.
Rev.  Lett. \us {67}, 2946 (1991).}

\REF\lan{L. D. Landau and I. Ya. Pomeranchuk, Dokl. Akad. Nauk. SSSR
92 (1953), 535, 735;
I. Ya. Pomeranchuk and E. L. Feinberg, Dokl. Akad. Nauk. SSSR 93
(1953), 439;
A.  B.  Migdal, Phys.  Rev \us{103}, 1811 (1956)}

In high energy inelastic hadron -- nucleus ($hA$) collisions, the
projectile rarely retains a major fraction of its momentum after
traversing the nucleus\refmark\bart.  Rather, its momentum is shared by
several produced particles, which form a hadron jet in the forward
direction.  The classical description of this phenomena is that the
hadron projectile suffers multiple collisions and repeated energy loss in
the nucleus.  However, the quantum mechanical situation in QCD is much
more interesting.  It is convenient to decompose the wavefunction of the
incoming hadron of lab momentum $P$ in terms of its free quark and gluon
Fock states.  Each Fock state of invariant mass $\cal M$ then persists
for a time
$2 P/({\cal M}^2-M_h^2)$
which for large $P$ is long
compared to the transit time through the nucleus.\refmark\GW~
Due to time dilation,
constituents which are separated by
a typical distance of 1 fermi in
impact space have no time to communicate while in the nucleus.  Thus at
high energies the quark and gluon constituents of the hadron typically
interact independently of each other.  Each quark or gluon constituent
can lose a finite fraction of its energy in its first collision in the
target due to QCD bremsstrahlung.
However, since there is insufficient time to regenerate its self-field,
repeated collisions (of similar hardness) by the same parton in the
target do not significantly increase
its total energy loss.\refmark\lan\
For similar reasons,
final state
hadrons are formed only after the projectile constituents have left the
nucleus.  Thus the nuclear interactions of a high energy hadron can be
most simply described in terms of the individual interactions of its
quarks and gluons. It is only necessary to take into account
coherent interactions between constituents
for the rare Fock components having a small
transverse size.\refmark\coh

\REF\bh{S.  J.  Brodsky and P.  Hoyer, Phys.  Rev.  Lett. \us{63}, 1566
(1989).}

The uncorrelated interactions of a high energy hadron's constituents in
nuclear matter imply that the constituents will normally
hadronize independently
of each other outside the nucleus, giving rise to overlapping jets in the
forward direction.  The rare case where a single hadron $h'$ carries a
large fraction $x_F$ of the beam momentum thus
most likely occurs when $h'$ is
formed from a transversely compact Fock state which can retain its
coherence while traversing the nucleus.  As we showed in an earlier
paper\refmark\bh, if the $A$-dependence of the inclusive cross section on
nuclei $d\sigma/dx_F( h A \to h' X)$ is parametrized as $A^\alpha(x_F)$,
then this restriction to compact states implies a monotonic decrease of
$\alpha$ with $x_F$, which is consistent with the trend of the
data\refmark\bart.

\REF\emc{EMC Collaboration, A.  Arvidson \etal, Nucl.  Phys. \us{B246},
381 (1984); R.  Windmolders, Proc. of the 24.  Int.  Conf. on High Energy
Physics, (Munich 1988, Eds.  R.  Kotthaus and J.  H.  K\"uhn, Springer
1989), p. 267.}

\REF\dy{D.  M.  Alde \etal, Phys.  Rev.  Lett. \us{64}, 2479 (1990)}

The time dilation argument given above implies that the fractional
energy loss of a high energy quark or gluon which participates in a hard
collision will not depend on the size of the nuclear target.
In fact, there is convincing empirical
evidence that high energy quarks can penetrate even very heavy nuclei
with insignificant mean energy loss.
In deep inelastic muon
scattering\refmark\emc\ on heavy nuclei, the struck quark emerges from
the nucleus with close to the full energy $\nu$ transferred by the
virtual photon, provided $\nu > 20$ GeV.  Secondly, the production cross
section of high mass muon pairs by hadrons on nuclei appears to be
closely proportional to the atomic number $A$ of the target\refmark\dy.
According to the Drell -- Yan mechanism, this implies that the projectile
quark (or antiquark) carries its full initial energy even at the time of
its annihilation with an antiquark (or quark) at the back side of the
target nucleus.  Both the deep inelastic lepton scattering and the large
mass lepton pair production data are thus incompatible with a significant
mean energy loss of quarks in nuclear matter.  The same conclusion
follows more generally from the factorization theorems of perturbative
QCD: the structure functions of the
projectile hadron and the fragmentation functions of the final state
partons are unmodified at leading twist by the nuclear
target.

\REF\muon{W.  Busza,
Nucl. Phys. \us {A544} 49c (1992), proceedings of the Ninth
International Conference
on Ultra-Relativistic Nucleus-Nucleus Collisions:
Quark Matter '91,  Gatlinburg, TN (November,1991).
The E665 Collaboration data is also reported in
J. J. Ryan, Ph. D. Thesis, Massachusetts Institute of Technology (1990).}

\REF\BDH{V.  Del Duca, S.  J. Brodsky and P.  Hoyer,
Phys.  Rev.  \us{D46}, 931 (1992).}

Recently, it has been pointed out\refmark\muon\ that data on fast hadron
leptoproduction even for (almost) real and virtual photons show no
nuclear effect.  For large photon energies $\nu>100$ GeV the hadron
momentum spectra observed in the E665 experiment are essentially
independent of the size of the nuclear target.  Even more remarkably, the
hadron spectrum in the kinematic region where a strong absorption effect
is seen in the low $Q^2$ inelastic cross section $(x_{Bj}<.005, Q^2<1$
GeV$^2)$ is very similar to the spectrum observed in the non-shadowing
region $(x_{Bj}>0.03, Q^2>2$ GeV$^2)$.

\REF\SHAD{For discussions of nuclear structure function shadowing in QCD,
see, e.g., L. L. Frankfurt and M. I. Strikman
Nucl. Phys. \us {B316} 340, (1989);
A. H. Mueller and J.-W. Qiu
Nucl. Phys. \us {B268} 427, (1986);
S. J. Brodsky and H. J. Lu, Phys. Rev. Lett. \us {64}, 1342, (1990);
V. Barone, M. Genovese, N. N. Nikolaev, E. Predazzi and B. G. Zakharov,
University of Torino preprint DFTT 14/92.}

We can
understand also
these effects in terms of the time
dilation and the QCD Fock state decomposition.
Clearly a major fraction
of the real or low $Q^2$ photoabsorption cross section occurs via
vector meson dominance.  However,
aside from diffractive processes,
the hadrons produced by the
intermediate vector mesons mostly have small
longitudinal momentum fractions,
just as in
hadron-nucleus scattering. By triggering on fast hadrons in the
photoabsorption cross section, we essentially eliminate the VMD
component and select the mechanism whereby the photon scatters only via
the $q\bar q$ intermediate Fock state.
As is well-known (see, \eg,
Ref. \BDH), the $q\bar q$ state scatters most strongly in an asymmetric
configuration where one of the quarks carries only a small fraction
$1-x\ll 1$
of the photon energy. The fluctuation
of the high energy photon to the $q \bar q$ state occurs at a time
$\tau\simeq 2\nu/({\cal M}^2 + Q^2),$
well before interactions occur in the target.
The transverse size $r_\perp$ of the $q\bar q$ Fock state on arrival
at the target nucleus can be estimated from the transverse velocity
$v_\perp=p_\perp/(1-x)\nu$ of the soft quark. In the low $Q^2$ region,
for $\M^2 \simeq p_\perp^2/(1-x)\gg Q^2$, we get a large size $r_\perp
= v_\perp \tau \simeq 2/p_\perp$, of the order of 1 fm for typical
values of $p_\perp$. Hence a soft interaction will occur on the nuclear
surface and the inner parts of the nucleus are shadowed.\refmark\SHAD~
For large
$Q^2\gg \M^2$ on the other hand, $r_\perp p_\perp \simeq 2\M^2/Q^2$ is
small, and there is no shadowing. Note that in {\it either case}
the major fraction $x$ of the photon momentum is carried by the
fast quark, which has no time to build up a self-field and thus loses
little energy in the nucleus. The fast hadrons, which are formed by
the fragmentation of this fast quark outside the nucleus, will thus have
a momentum spectrum which is independent of the nuclear size both in the
(shadowing) region of small $Q^2$ and in the (transparent) region
of large $Q^2$.

\REF\bad{J.  Badier \etal, Z.  Phys. \us{C20}, 101 (1983).}

\REF\kat{S.  Katsanevas \etal, Phys.  Rev.  Lett. \us{60}, 2121 (1988).}

\REF\ald{D.  M.  Alde \etal, Phys.  Rev.  Lett. \us{66}, 2285 (1991); \us
{66}, 133 (1991).} \REF\hsv{P.  Hoyer, M.  V\"anttinen and U.  Sukhatme, Phys.
Lett. \us {B246} 217 (1990).}

\REF\BM{S.  J.  Brodsky and A.  H.  Mueller, Phys.  Lett. \us {206B}, 685
(1988).}

\REF\HUFNER{%
For an illustration of this effect, see J. Hufner and M. Simbel
Phys. Lett. \us {B258}, 465 (1991);
J. Hufner, B. Povh, and S. Gardner
Phys. Lett. \us {B238}, 103 (1990).}

The data\refmark{\bad,\kat,\ald}~ for the hadroproduction of heavy
quarkonium states on nuclei show a strong nuclear suppression at large
$x_F$,
which is in striking contrast to the minimal effects seen in
continuum lepton pair production.  Since the nuclear dependence does not
factorize\refmark\hsv~ as a function of the nuclear parton fraction
$x_2$, the effect cannot be due to gluon shadowing.  The breakdown of
factorization also implies that the nuclear dependence must be associated
with a higher twist mechanism.  Furthermore, the nuclear suppression seen
in the E772 experiment\refmark\ald~ is essentially identical for $J/\psi,\
\psi',$ and $\psi''$ production even though these states have drastically
different sizes; thus the nuclear effect cannot be attributed to final
state hadronic absorption.  In fact, at high $x_F$ the $c \bar c$
pair does not form
the quarkonium state until it is well beyond the nuclear
volume.\refmark\BM\
Thus final state absorption of heavy quarkonium is predicted to
decrease with growing $x_F,$
contrary to the observed nuclear attenuation.\refmark\HUFNER

\REF\gm{S.  Gavin and J.  Milana, Phys.  Rev.  Lett. \us{68}, 1834
(1992).}

\REF\qu{E.  Quack, Heidelberg preprint HD-TVP-92-2 (June 1992).}

\REF\ff{S. Frankel and  W. Frati, University of Pennsylvania preprint
UPR-0499T (May 1992).}

Recently, some authors have claimed\refmark{\gm,\qu,\ff}
that the anomalous
suppression of large $x_F$ $J/\psi$ production on nuclear
targets\refmark{\bad,\kat,\ald} can be explained by postulating a
significant
energy loss for fast gluons and quarks as they
propagate through the nucleus.  The nuclear effect is assumed to be
higher twist so that it would not conflict with the PQCD factorization
theorems.  Any parton energy loss
implies
that the structure function of the projectile is sampled at a larger
value of $x_1$ than would otherwise be inferred from the $x_F$ of the
$J/\psi$.

To see the effect of such a parton energy loss explicitly, let us assume
that the structure function has a behavior $F(x_1)\propto (1-x_1)^n$.
The suppression corresponding to a fractional energy loss $\Delta x_1$ is
$${F(x_1+\Delta x_1)\over F(x_1)} \simeq 1-{n\over 1-x_1}\Delta x_1
\simeq A^{\delta\alpha}. \eqn\sup$$
Hence the effective shift
$\delta\alpha$ in the nuclear power dependence is approximately given by
$$ \delta\alpha = - {n\over 1-x_1}{\Delta x_1\over\log A}. \eqn\dela$$
An energy loss due to multiple scattering in the nucleus would be
proportional to the nuclear diameter, $\Delta x_1 \propto A^{1/3}$.  Then
the dependence of $\delta\alpha$ on $A$ in Eq. \dela\ is indeed quite
weak; $A^{1/3}/\log A =1$ within 10\% over the range $5\leq A\leq 200$,
implying that the energy loss effect can be indeed parametrized as a
power of $A$.

The authors of Refs. \gm~ and \qu~ assume that the average fractional
momentum loss of an incident parton in a high $Q^2$ reaction, such as
charm production, is given by
$${\Delta E_{lab}\over{E_{lab}}}={\Delta x_1=C {x_1\over Q^2}}
 A^{1/3}\eqn\gami$$
where $C$ is a color-dependent constant and
$A^{1/3}$ reflects the number of nuclear collisions.  They propose that
the coefficient of $A^{1/3}$ decreases as $1/Q^2$ because energy loss
should be a higher twist effect; it is also evidently consistent with the
reduced nuclear suppression observed for the $\Upsilon$ data.  Finally,
they argue that the fractional loss should be proportional to $x_1$ in
analogy with the QCD bremsstrahlung processes.

\REF\QM{A report on this work was given in S.  J.  Brodsky,
Nucl. Phys. \us {A544} 223c (1992), proceedings of the Ninth
International Conference
on Ultra-Relativistic Nucleus-Nucleus Collisions:
Quark Matter '91,  Gatlinburg, TN (November,
1991), edited by T. C. Awes, et al.
The fact that the  energy loss is constant in the target rest system
has also been discussed by A.  H.  Mueller at the
Penn State Workshop on Nuclear Effects in QCD,
(March, 1992, unpublished).}

Here we would like to show that the form \gami\ of the energy loss
violates general quantum mechanical arguments based on the uncertainty
principle.  We shall show that any $A$-dependent energy loss $\Delta
E_{lab}$ must be independent of $E_{lab}$.  Thus the energy loss per unit
length of the target is fixed\refmark\QM~ in the target rest frame,
rather than being $\propto E_{lab}$ as in Eq. \gami.  In fact, the energy
loss due to multiple soft scattering cannot depend on the $Q^2$ of the
hard collision: it occurs long before the hard vertex and is thus
causally independent of $Q^2$.  Furthermore, since $Q^2=sx_1x_2$, Eq.
\gami\ would imply that the multiple scattering energy loss depends on
the $x_2$ of the target parton involved in the hard collision.

\REF\bbl{G.  Bodwin, S.  J.  Brodsky, and G.  P.  Lepage, Phys.  Rev.
{\us D39}, 3287 (1989)}

The requirement of a fixed, energy-independent energy loss in the target
rest frame is a direct consequence of the uncertainty principle relation
$\Delta p_z \Delta L > 1$.  The uncertainty principle sets a minimum
longitudinal momentum transfer $\Delta p_z$ from the target for any
inelastic process which can be resolved as occurring between two
scattering centers of separation $L$.  The longitudinal momentum transfer
to the scattered parton due to gluon radiation is $\Delta p_z
\simeq\Delta {\cal M}^2/2 E_{lab}$, where $\Delta {\cal M}^2\sim
k^2_{\perp g}/x_g$ is the difference between the incident parton mass
squared and the mass squared of the parton--gluon system after radiation,
and $k_{\perp g}$, $x_g$ are, respectively, the transverse momentum and
momentum fraction of the radiated gluon.  Repeated radiation at distances
less than that allowed by the uncertainty relation is cancelled because
of destructive interference between the radiation emitted by the parton
at the two scattering centers\refmark\bbl.  The minimum distance $L$
required between repeated emissions may be interpreted as that needed for
the buildup of the gluon field around the bare parton.

For a simple and explicit example of how the uncertainty principle is
upheld in perturbation theory, consider the photon radiation induced by
the double scattering of a scalar electron (Fig. 1).  We shall keep the
times $t_1,t_2$ of the instantaneous Coulomb exchanges fixed --- they
represent two interactions in the same nucleus.  The photon can be
radiated at a time $t$ before, in between, or after the Coulomb
scatterings, corresponding to the three diagrams illustrated in Fig. 1.
Up to a common factor (which includes the Coulomb propagators), the
amplitudes of the three time orderings are
$$\eqalign{ M_a(t<t_1,t_2)=&-i
\exp[-iE_{a2}(t_2-t_1)]
\int_{-\infty}^{t_1}dt\,\exp[-i(E_{a1}-E)(t_1-t)]\cr
=&{\vec\varepsilon\cdot\vec p\over{E-E_{a1}}}\exp(-iE_{a2}\Delta t)\crr
M_b(t_1<t<t_2)=&{\vec\varepsilon\cdot(\vec p+\vec\ell_1)\over
{E_{b1}-E_{b2}}}[\exp(-iE_{b1}\Delta t)-\exp(-iE_{b2}\Delta t)]\crr
M_c(t_1,t_2<t)=&{\vec\varepsilon\cdot(\vec p+\vec\ell_1+\vec\ell_2)\over
 {E-E_{c2}}}\exp(-iE_{c1}\Delta t)\cr} \eqn\tfix$$
Here $\Delta t=
t_2-t_1$ and $E_{a1},\ldots,E_{c2}$
are the energies of the scattering
system at the intermediate times indicated in Fig. 1.

As the initial (and final) scattering energy grows, ($E\to\infty$ at
fixed fractional momentum $x_\gamma=k_\parallel/E$
of the photon), all of the
intermediate energies approach E.  For example,
$$E-E_{a1}\simeq
-{1\over{2E(1-x_\gamma)}}(x_\gamma m^2+{k_\perp^2\over x_\gamma}).
\eqn\ediff $$
Thus at fixed\footnote*{Or for any $\Delta t \lsim L_A$, as would be the
case for scattering in a nucleus of diameter $L_A$.} $\Delta t$ all phase
factors in Eq. \tfix\ approach $\exp(-iE\Delta t)$.  The amplitudes $M_a$
and $M_c$ then each have the same form as the amplitude for a single
Coulomb scattering with momentum exchange
$\vec\ell=\vec\ell_1+\vec\ell_2$.  The amplitude $M_b$, which describes
photon emission between the two Coulomb exchanges, is of ${\cal O}(1/E)$
compared to $M_a+M_b$, due to the cancellation of the phase factors in
brackets.  Hence the double scattering is not resolved, and the strength
of the single scattering is renormalized.  This is precisely the content
of the uncertainty relation, stating that multiple scattering in a target
of fixed length cannot induce fractional energy loss in the high energy
limit.

On the other hand, we can also see from Eq. \ediff\ that $M_b$ is of the
same order as $M_{a,b}$ if the photon momentum fraction $x_\gamma={\cal
O}(1/E)$.  Hence multiple scattering can induce a {\it fixed} energy loss
in the laboratory frame.  In general, the fractional energy loss
$x_\gamma$ that can be induced by multiple scattering in a target of
length $L_A$ is limited by
$$x_\gamma < {k^2_\perp L_A \over 2 E}
\eqn\xg$$
where $k_\perp$ is the transverse momentum of the photon.

As discussed above, the same bound \xg\ can be obtained directly from the
uncertainty relation, and thus applies equally to gluon radiation by
incoming or outgoing partons in hadron scattering.  Hence the bound on
the fractional energy loss $\Delta x_1$ of the projectile parton
appearing in Eqs. \sup,\dela\ is given by
$$\Delta x_1 \lsim {\kappa\over  x_1 s} A^{1/3}. \eqn\deltax$$
where we used $E = x_1 s/2 M_P$ and took
the nuclear radius $R_A \sim 1.2~{\rm fm} A^{1/3}$ to characterize the
largest effective distance between scattering centers in the nucleus.
Hence
$$\kappa \sim (1.2~{\rm fm}) M_p <k^2_\perp> \sim 0.5 ~GeV^2.
\eqn\kap$$
since gluons radiated by the incident or final state partons
in cold nuclear matter have a characteristic transverse momentum
$<k^2_\perp> \sim 0.1~{\rm GeV}^2.$

The bound \deltax\ should be contrasted with the assumption \gami\ of
Refs. \gm~ and \qu.  Our bound is independent of $Q^2$, since the range
of the hard interaction is short and does not affect multiple scattering
elsewhere in the nucleus.\footnote*{Gluon radiation may occur in the hard
process itself even at the leading twist level, as in $gg\to c\bar cg$.
These processes are not suppressed at high $Q^2$ and are given by the
higher order perturbative terms in the hard cross-section $\hat\sigma$.}
The bound \deltax\ is also independent of the color charge of the parton,
\ie, this upper bound is the same for quarks, gluons and compact $c\bar
c$ states.  Most importantly, the bound \deltax\ is inversely
proportional to the laboratory energy of the radiating parton.  Hence
energy loss becomes insignificant at high energies, and the
cross-sections obey Feynman scaling.  The fact that the measured
cross-sections\refmark{\bad,\ald} indeed satisfy Feynman scaling shows
that the effects of finite energy loss is already negligible in the data
for $E_{lab} \gsim 100$ GeV.

The bound \deltax\ implies numerically insignificant effects of energy
loss in the high energy data.  Consider, for example, the suppression of
$J/\psi$ production on Tungsten, which for 800 GeV protons was measured
by the E772 Collaboration\refmark\ald\ to be 60\% at $x_F=0.64$.
Substituting \deltax\ into \dela\ gives a suppression due to parton
energy loss of
$\delta \alpha = - 0.008$ for $n=5$,
corresponding to only a 4\%
suppression of the cross section on Tungsten. This is negligible
compared to the effect seen in the data.

\REF\gyu{M.  Gyulassy, M.  Plumer, M.  Thoma and X.  N.  Wang, LBL-31002,
(1991); X.~N.~Wang, private communication.}

\REF\tho{M.H.  Thoma and M.  Gyulassy, Nucl.  Phys. \us{B351}, 491
(1991).}

\REF\intr{S.~J.  Brodsky, P.~Hoyer, C.~Peterson, and N.~ Sakai, Phys.
Lett. \us{93B}, 451 (1980); S.~J.  Brodsky, C.~Peterson, and N.~ Sakai,
Phys.  Rev. \us{D23}, 2745 (1981).}

\REF\bhmt{S.  J.  Brodsky, P.  Hoyer, A.  H.  Mueller and W.-K.  Tang,
Nucl.  Phys.  \us{B369}, 519 (1992). For a detailed comparison with
data, see R. Vogt, S. J. Brodsky, and P. Hoyer,
Nucl. Phys. \us{B360} 67, (1991).}

According to \xg, the average radiative energy loss per unit distance in
nuclear matter is $dE/dz \lsim
{1 \over 2}<k^2_\perp> \simeq 0.25$ GeV/fm.  A
similar degradation of energy is expected from elastic
scattering\refmark\gyu.  The total expected energy loss, $dE/dz \sim 0.5$
GeV/fm, appears to be consistent with an estimate using combined SLAC and
EMC data for jet fragmentation in nuclei\refmark\gyu.  At high energies,
such a fixed energy loss becomes insignificant, thus explaining the lack
of nuclear target dependence of jet fragmentation processes in deep
inelastic lepton scattering\refmark\emc~ and the lack of nuclear-induced
initial state energy loss of the annihilating quark in massive lepton
pair production\refmark\dy. On the other hand, in a medium which is at
high temperature $T$, such as in a quark gluon plasma, the average
energy loss can be larger\refmark{\gyu,\tho}, since
$<k_\perp>\propto T$.

The nuclear suppression of quarkonium production at high $x_F$ is
observed to be mass-dependent --- the suppression measured in the E772
experiment is smaller for $\Upsilon$ production than for $J/\psi$
production\refmark\ald.  In view of the $Q^2$-independence of Eq.
\deltax~ at fixed $s$ and $x_1$, this again rules out an explanation of
the $A$-dependence in terms of nuclear-induced energy loss.  It also
should be emphasized that the $J/\psi$ cross-section at large $x_F$ is
measured\refmark\bad\ to be {\it in excess} of that predicted by leading
twist fusion processes for proton targets.
Thus it is likely
that the anomalous nuclear effects are due to higher twist effects
which enhance the hard cross section on elementary targets.

Sizeable
higher twist contributions at large $x_F$ are in fact expected
in QCD from intrinsic heavy quark production mechanisms\refmark{\intr}.
In contrast to the leading twist fusion contributions such as $g g \to c
\bar c$, the intrinsic contributions involve two or more constituents in
the projectile.  Although these amplitudes are relatively suppressed by
powers of $1/m^2_{Q \bar Q},$ a greater fraction of the projectile's
momentum is involved so they are less suppressed at high $x_F.$ Since the
slow spectators interact in the target, the intrinsic contributions to
the large $x_F$ cross section have nuclear dependence close to
$A^{2/3}$\refmark\bh.  Recently, it was shown that the above qualitative
features of intrinsic production emerge in a perturbative QCD
analysis\refmark\bhmt.  However, a definitive
explanation of the nuclear
anomalies in heavy quark production at large $x_F$ in terms of higher
twist contributions will require a more quantitative analysis of
multiparton correlations in QCD.

\refout
\endpage
FIGURE CAPTION

\noindent
Fig. 1. Photon radiation diagrams associated with the double Coulomb scattering
of a (scalar) electron at the fixed times $t_1$ and $t_2$. The total initial
and
final energy of the scattering is $E$, and the intermediate energies at the
times indicated in (a), (b) and (c) are denoted by $E_{ai}$, $E_{bi}$ and
$E_{ci}$ $(i=1,2)$, respectively.

\end